# GAGA: A PARSIMONIOUS AND FLEXIBLE MODEL FOR DIFFERENTIAL EXPRESSION ANALYSIS

By David Rossell

*Institute for Research in Biomedicine of Barcelona*

Hierarchical models are a powerful tool for high-throughput data with a small to moderate number of replicates, as they allow sharing information across units of information, for example, genes. We propose two such models and show its increased sensitivity in microarray differential expression applications. We build on the gamma–gamma hierarchical model introduced by Kendziorski et al. [*Statist. Med.* **22** (2003) 3899–3914] and Newton et al. [*Biostatistics* **5** (2004) 155–176], by addressing important limitations that may have hampered its performance and its more widespread use. The models parsimoniously describe the expression of thousands of genes with a small number of hyper-parameters. This makes them easy to interpret and analytically tractable. The first model is a simple extension that improves the fit substantially with almost no increase in complexity. We propose a second extension that uses a mixture of gamma distributions to further improve the fit, at the expense of increased computational burden. We derive several approximations that significantly reduce the computational cost. We find that our models outperform the original formulation of the model, as well as some other popular methods for differential expression analysis. The improved performance is specially noticeable for the small sample sizes commonly encountered in high-throughput experiments. Our methods are implemented in the freely available Bioconductor `gaga` package.

**1. Introduction.** A main challenge in microarray and other high-throughput experiments is the limited number of replicated measurements that are obtained for each gene. That is, data is abundant at an overall level but it is scarce at the gene level, and, therefore, there is much potential for methods that allow for the sharing of information across genes. This feature is particularly important due to the small sample sizes that are frequently encountered in these studies.

Hierarchical models naturally allow for this kind of information sharing. Typical examples are Lönnstedt and Speed (2002) and Smyth (2004), who modeled gene-specific parameter estimates via hierarchical empirical Bayes methods to obtain improved testing procedures. Kendziorski et al. (2003), Newton et al. (2001) and Newton and Kendziorski (2003) proposed hierarchical models that depend on few hyper-parameters, that is, they greatly reduce the dimensionality of the problem.

We propose a novel approach for massive multiple inference. From here on we focus on the analysis of differential expression in microarrays, although the approach can be used for other forms of high-throughput data as well. The proposed model builds on the gamma–gamma hierarchical model of Kendziorski et al. (2003). This model is methodologically and mathematically attractive, but has only had a modest effect in the practice of expression analysis. We identify some data analysis issues that might be limiting factors to prevent a more widespread application of the gamma–gamma model, and propose an improved model to address this issues. Our approach combines features from Lo and Gottardo (2007) by specifying varying coefficients of variation across genes, and Yuan and Kendziorski (2006) by specifying a mixture prior on gene-specific parameters that induces gene clustering.

In particular, the gamma–gamma model assumes that the observations for each gene arise from a gamma distribution with common shape parameter across all genes and a scale parameter that arises from a hierarchical gamma prior. Since the model uses a single gamma prior, we refer to it as the *Ga model*. We find the gamma choice appealing, for it is a flexible family that can capture the asymmetric distributional shapes frequently encountered in microarrays, even after log-transforming. As we show in this paper, although the Ga model is elegant and parsimonious, it fails to provide an adequate fit in a number of examples. The main challenge in adding flexibility to the model is that it seriously complicates the computations required for model fit and inference. We propose an extension in two directions. First, we specify a gamma prior on both the shape and the inverse mean parameters (*GaGa model*). The extension is still parsimonious, requiring only one additional hyper-parameter, and it can be fit in a computationally efficient manner. We develop an algorithm that requires a computational effort comparable to the Ga model. In a second extension we specify a gamma prior on the shape parameter and a mixture of gamma priors on the inverse mean (*MiGaGa model*). This provides additional flexibility, albeit at the expense of reduced model parsimony and increased computational cost.

In summary, the aim of this paper is to improve differential expression analysis by providing a method with higher sensitivity than several standard approaches. This is achieved by extending the basic Ga model while maintaining its methodological beauty and closed-form inference. The extension addresses data analytic issues of practical relevance, and which may have prevented the more widespread use of the model.

The paper is structured as follows. In Section 2 we review the Ga model and we present its first extension: the GaGa model. We derive expressions for posterior probabilities of interest, and point out several schemes to fit the model. In Section 3 we propose as a further generalization the MiGaGa. For both extensions, the posterior distributions of the gamma shape parameters are known only up to a constant. We refer to this distribution, which to our knowledge has not been described before, as the *gamma shape distribution*. In Section 4 we derive useful approximations for this distribution. In Section 5 we outline how to find differentially expressed (DE) genes, and in Section 6 we apply our approach to simulated data and several examples. Some concluding remarks follow in Section 7. The methods described in this paper are implemented in the R package gaga, available as part of Bioconductor 2.2.

**2. The GaGa model.** We assume that the data has been background corrected, normalized and quantified in a sensible manner [Dudoit et al. (2002); Stafford (2008)]. Let $x_{ij}$ be the measure of expression for gene $i$, $i = 1, \ldots, n$, in microarray $j$, $j = 1, \ldots, J$. Let $z_j \in \{1, \ldots, K\}$ indicate group membership, for example, $z_j = 1$ for normal cells and $z_j = 2$ for cancer cells. We denote the vector of observations for gene $i$ as $\mathbf{x}_i$ and the whole data as $\mathbf{x}$. We use $\text{Ga}(\cdot)$ to denote a gamma distribution, $\text{IGa}(\cdot)$ for the inverse gamma, $\text{Mult}(\cdot)$ for the multinomial, $\text{Dirichlet}(\cdot)$ for the Dirichlet and $\text{GaS}(\cdot)$ for the gamma shape distribution. The GaS distribution arises as the posterior distribution of the gamma shape parameter conditional on the observed data and on some other model parameters. To our knowledge it has not been introduced before, so we present its definition in Section 4.

In the differential expression problem the investigator is interested in determining the expression pattern that each gene follows. This inference problem can be viewed as a hypothesis testing problem. Throughout we use the terms *hypothesis* and *expression pattern* interchangeably. The simplest setup is having $K = 2$ groups and 2 hypotheses: pattern 0 under which both groups are equally expressed (null hypothesis) and pattern 1 under which they are differentially expressed (alternative hypothesis). For $K > 2$ we may want to consider more than 2 patterns. For example, if group 1 corresponds to normal cells, group 2 to cells with type A cancer and group 3 to type B cancer, one may be interested in assigning each gene to one of the following patterns:

(1)
Pattern 0: Normal = Cancer A = Cancer B,

Pattern 1: Normal ≠ Cancer A = Cancer B,

Pattern 2: Normal ≠ Cancer A ≠ Cancer B.

Denote by $H$ the number of hypotheses, and let the latent variable $\delta_i \in \{0, 1, \ldots, H-1\}$ indicate the true expression pattern for gene $i$. We refer to genes with $\delta_i = 0$ as equally expressed (EE) and genes with $\delta_i \neq 0$ as differentially expressed (DE), and we denote $\boldsymbol{\delta} = (\delta_1, \ldots, \delta_n)$.

2.1. *The model.* The Ga model [Kendziorski et al. (2003); Newton et al. (2001); Newton and Kendziorski (2003)] assumes that $x_{ij}$ are independent realizations from $\text{Ga}(\alpha_i, \lambda_{i,z_j})$ (i.e., the mean is $\alpha_i/\lambda_{i,z_j}$). The model assumes $\delta_i \sim \text{Mult}(1, \boldsymbol{\pi})$, fixes $\alpha_i = \alpha$ for all $i$ and specifies the hierarchical prior $\lambda_{i,z_j} \sim \text{Ga}(\alpha_0, \nu)$ for all distinct scale parameters under pattern $\delta_i$. Here $(\alpha_0, \nu, \alpha, \boldsymbol{\pi})$ are hyper-parameters common to all genes. For $\delta_i = 0$ (EE genes) we have $\lambda_{i,1} = \cdots = \lambda_{i,K}$, and for $\delta_i \neq 0$ some of the $\lambda_{i,z_j}$ are different from each other, according to the specification of the hypotheses.

The Ga model imposes the restriction that $1/\sqrt{\alpha_i}$, the within-groups coefficients of variation (CV), must be constant across all genes and groups. The assumption is analytically convenient, but we have found it not to be reasonable in typical data. Figure 1(a) shows empirical CVs for the Armstrong data, described in Section 6. The sample CVs and mean expressions are roughly independent. This does not confirm, however, the constant CV assumption. If the true CVs were indeed constant, the sample CVs should be similar to each other (up to estimation error), whereas we observe CVs that range from 0.005 to 0.7. We conducted a simulation study and confirmed that, under the constant CV assumption, the range of sample CVs should be much smaller. To assess the extent to which this lack of fit affects the inference, Figure 1(a) highlights the genes de-clared DE by the Ga model. These are mostly genes with above average CV. This is due, we believe, to the constant CV assumption, which makes the Ga model view atypical CVs as evidence for differential expression.

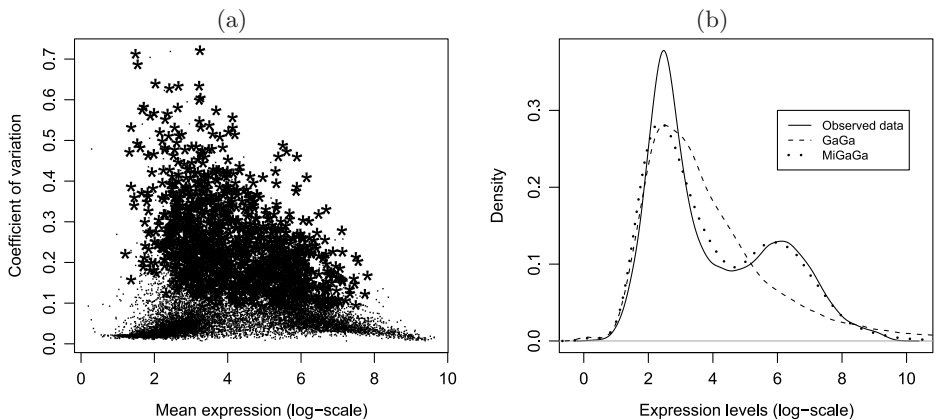

FIG. 1. *Armstrong data.* (a) *Sample mean and CV for each gene (* denotes genes declared DE by the Ga model).* (b) *Marginal distribution of data (*$\log_2$ *scale vs. prior predictive of GaGa and MiGaGa with $M = 2$ and $\boldsymbol{\omega} = \hat{\boldsymbol{\omega}}$).*

In practice an analyst would certainly interrogate the identified genes and down-weight cases like those identified in Figure 1(a). Still, it is desirable to allow $\alpha_i$ in the model to depend on the gene $i$ and automate this step when possible. The main difficulty with this generalization is that it seriously complicates computations. We propose a generalization that addresses this limitation in a computationally efficient manner. We introduce gene-specific shape parameters $\alpha_i$ and assume $x_{ij} \sim \text{Ga}(\alpha_i, \alpha_i/\lambda_{i,z_j})$ (i.e., $\lambda_{i,z_j}$ is the mean), with the following hierarchical prior:

$$
\begin{aligned}
\lambda_{i,k}|\delta_i, \alpha_0, \nu &\sim \text{IGa}(\alpha_0, \alpha_0/\nu), \qquad \text{indep. for } i=1,\ldots,n \\
\alpha_i|\delta_i, \beta, \mu &\sim \text{Ga}(\beta, \beta/\mu), \qquad \text{indep. for } i=1,\ldots,n,
\end{aligned}
\tag{2}
$$

and a prior for $\delta_i$ as before. We refer to (2) as the GaGa model. As in the Ga model, when $\delta_i = 0$ we have $\lambda_{i1} = \cdots = \lambda_{iK}$, whereas under $\delta_i \neq 0$ some of them are different from each other (although they still arise from the same marginal distribution). The GaGa model replaces the hyper-parameter $\alpha$ of the Ga model with the pair $(\beta, \mu)$. That is, the additional flexibility is achieved with only one more hyper-parameter.

Note that we assume that $\alpha_i$ is constant across groups. Allowing it to vary across groups would allow to compare not only mean expression levels but the full distribution, which is biologically relevant (Lapointe et al. (2004); Coombes, Wang and Baggerly (2007)). Even though in this paper we focus on the comparison of $\lambda_{i,z_j}$, we have also derived and implemented the comparison of $\alpha_i$ in our Bioconductor package gaga.

2.2. *Posterior distributions.* We derive the posterior probability that a gene follows each expression pattern, which is needed to obtain lists of differentially expressed genes. We also present the posterior distribution of the first-stage parameters, which are needed to obtain fold change estimates. Both posterior probabilities and distributions are derived assuming that the hyper-parameters $\boldsymbol{\omega} = (\alpha_0, \nu, \beta, \mu, \boldsymbol{\pi})$ are fixed, as was done for the Ga model. We denote the vector of means for a single gene as $\boldsymbol{\lambda}_i = (\lambda_{i,1}, \ldots, \lambda_{i,K})$ and we let $\boldsymbol{\lambda} = (\boldsymbol{\lambda}_1, \ldots, \boldsymbol{\lambda}_n)$, $\boldsymbol{\alpha} = (\alpha_1, \ldots, \alpha_n)$ be the collection of these parameters. From (2) we see that, conditional on $\boldsymbol{\omega}$, the gene-specific parameters $(\delta_i, \alpha_i, \boldsymbol{\lambda}_i)$ are independent *a posteriori* across genes $i = 1, \ldots, n$. Therefore, it suffices to derive the posterior for each gene separately.

Let $P_i$ be the log-product of the observations for gene $i$, that is, $P_i = \log \prod_{j=1}^{J} x_{ij}$, and let $N_{\delta_i}$ be the number of groups that are distinct under pattern $\delta_i$. In our example in (1) we have $H = 3$ patterns: under pattern 0 we have $N_0 = 1$ distinct groups, and, similarly, $N_1 = 2$, $N_2 = 3$. Let $S_{i,\delta_i,k}$ for $i = 1, \ldots, n$, $\delta_i = 0, \ldots, H-1$ and $k = 1, \ldots, N_{\delta_i}$ be the sum of observations from gene $i$ that under pattern $\delta_i$ correspond to the $k$th distinct group and $J_{i,\delta_i,k}$ be the number of terms in the sum. In our example $S_{10,0,1}$ denotes the sum of all observations from gene 10 (since under pattern 0 there is only one distinct group), $S_{10,1,1}$ denotes the sum of observations from normal samples (since it is the first distinct group under pattern 1) and $S_{10,1,2}$ the sum from cancers of type A and B. We denote by $\mathbf{S}_{i,\delta_i}$ and $\mathbf{J}_{i,\delta_i}$ the collection of $S_{i,\delta_i,k}$ and $J_{i,\delta_i,k}$ for $k = 1, \ldots, N_{\delta_i}$.

The posterior probability that gene $i$ follows expression pattern $l$, which we denote as $v_{il}$, is given by $v_{il} = P(\delta_i = l|\mathbf{x}, \boldsymbol{\omega}) \propto f(\mathbf{x}_i|\delta_i = l, \boldsymbol{\omega})\pi_l$ for $l = 0, \ldots, H-1$, where $f(\mathbf{x}_i|\delta_i, \boldsymbol{\omega})$ is equal to

$$(3) \quad \left[\frac{(\alpha_0/\nu)^{\alpha_0}(\beta/\mu)^\beta}{\Gamma(\alpha_0)\Gamma(\beta)}\right]^{N_{\delta_i}} \prod_{k=1}^{N_{\delta_i}} \frac{1}{C(\mathbf{J}_{i,\delta_i}, \beta, \beta/\mu - P_i, \alpha_0, \alpha_0/\nu, \mathbf{S}_{i,\delta_i})},$$

and $C(\cdot)$ is the gamma shape normalization constant, defined in Section 4.

The distribution of $(\alpha_{ik}, \lambda_{ik})$ conditional on the observed data, $\delta_i$ and the hyper-parameters $\boldsymbol{\omega}$ is

$$(4) \quad \begin{aligned} \alpha_i|\delta_i, \boldsymbol{\omega}, \mathbf{x} &\sim \mathrm{GaS}(\mathbf{J}_{i,\delta_i}, \beta, \beta/\mu - P_i, \alpha_0, \alpha_0/\nu, \mathbf{S}_{i,\delta_i}), \\ \lambda_{i,k}|\alpha_i, \delta_i, \boldsymbol{\omega}, \mathbf{x} &\sim \mathrm{IGa}(\alpha_i J_{i,\delta_i,k} + \alpha_0, \alpha_0/\nu + \alpha_i S_{i,\delta_i,k}). \end{aligned}$$

For any given $\boldsymbol{\omega}$, (4) can be used to obtain posterior credibility intervals in the usual fashion. Note that $\alpha_i$ and $\lambda_{i,k}$ are not conditionally independent *a posteriori* given $(\delta_i, \boldsymbol{\omega})$ as they are a priori.

2.3. *Model fitting.* The model can be fit by estimating $(\alpha_0, \nu, \beta, \mu, \boldsymbol{\pi})$ with an empirical Bayes argument. To this end, we implemented an expectation-maximization algorithm completely analogous to that for the Ga model [Kendziorski et al. (2003), Appendix]. The EM algorithm is described in the Supplementary Material [Rossell (2009)]. Alternatively, fully Bayesian model fitting schemes which specify a hyper-prior on $(\alpha_0, \nu, \beta, \mu, \boldsymbol{\pi})$ are also possible and are provided in the Supplementary Material [Rossell (2009)]. Both EM and fully Bayesian approaches are implemented in our `gaga` package, and they usually deliver virtually identical results. This is to be expected, as microarray data is strongly informative about parameters that are common to all genes.

Both the empirical and fully Bayesian algorithms require the evaluation of the normalization constants in the posterior distribution of the gamma shape parameters $\alpha_1, \ldots, \alpha_n$, which are not available in closed form. In Section 4 we derive useful approximations that reduce the computational burden.

**3. The MiGaGa model.** The GaGa model addresses the problem illustrated in Figure 1(a) by allowing varying CVs across genes. However, another limitation remains. In practice, some normalization procedures, such as RMA [Irizarry et al. (2003)] or GCRMA [Wu et al. (2004)], oftentimes result in a distinctly bimodal distribution for the gene expressions. Figure 1(b) shows a kernel density estimate of the density of $x_{ij}$ for the Armstrong data (see Section 6.1), and compares it with the prior-predictive under the GaGa model. The model does not capture the bimodality. To address this limitation, we introduce a further generalization, by letting $\lambda_{i,k}$ arise from the mixture

$$\lambda_{i,k}|\delta_i, \boldsymbol{\rho}, \boldsymbol{\alpha}_0, \boldsymbol{\nu} \sim \sum_{m=0}^{M} \rho_m \, \mathrm{IGa}(\alpha_{0m}, \alpha_{0m}/\nu_m),$$

(5)
$$\boldsymbol{\rho} \sim \mathrm{Dirichlet}(\mathbf{r}),$$

where $M$ is the number of components in the mixture. The rest of the model is as in (2). The posterior distributions and model fitting procedures are largely analogous to that for the GaGa model and are detailed in the Supplementary Material [Rossell (2009)]. The main difference is that for the MiGaGa one introduces latent variables indicating the cluster to which each gene belongs.

Compared to the GaGa prior, the additional flexibility in MiGaGa potentially allows us to obtain a better fit to the data, albeit this comes at the cost of increased model complexity and computational burden. Figure 1(b) shows how the MiGaGa prior predictive improves the GaGa fit substantially. We selected $M = 2$ clusters as they capture the bi-modality observed in Figure 1(b). In general, one can either select $M$ maximizing some criterion such as the BIC [Schwarz (1978)] or simply fit a MiGaGa model with large $M$. In the latter case, after the model fit one could remove the clusters with estimated weights $\rho_m$ below some threshold.

**4. The gamma shape distribution.** The posterior distribution of the shape parameter $\alpha_i$ in (2), which we refer to as *gamma shape distribution*, has not been described before. It is similar to the distribution that arises when the parametrization is in terms of the shape and scale parameters [Damsleth (1975); Miller (1980)]. To simplify notation, we denote by $y$ a positive continuous random variable that follows this distribution. Its probability density function, indexed by the parameters $\mathbf{a} = (a_1, \ldots, a_p)$ where $a_i \geq 0$, $b > 0$, $d > 0$, $r > 0$, $\mathbf{s} = (s_1, \ldots, s_p)$ where $s_i > 0$, $c > -\sum_{i=1}^{p} a_i \log(s_i/a_i)$, can be written as $f(y|\mathbf{a}, b, c, d, r, \mathbf{s}) =$

(6) $$C(\mathbf{a}, b, c, d, r, \mathbf{s}) y^{b-pd-1} e^{-yc} \prod_{i=1}^{p} \frac{\Gamma(a_i y + d)}{\Gamma(y)^{a_i}} \left(\frac{y}{r + s_i y}\right)^{a_i y + d}$$

for $y > 0$. $C(\mathbf{a}, b, c, d, r, \mathbf{s})$ is the normalization constant and $\Gamma(\cdot)$ is the gamma function. For $a_1 = \cdots = a_p = d = 0$, (6) simplifies to a gamma distribution.

In general, to obtain random draws from (6) or to evaluate $C(\mathbf{a}, b, c, d, r, \mathbf{s})$, one has to resort to numerical methods. This is impractical in our setup, since both the EM algorithm and the fully Bayesian fitting schemes [see Supplementary Material in Rossell (2009)] require performing these steps a very large number of times. Approximations are required to decrease the computational burden.

We start by deriving an approximation to (6) that is appropriate for large values of $y$. By approximating $\Gamma(\cdot)$ with Stirling's formula and evaluating the limit of the resulting expression as $y \to \infty$ and $a_i y + d \to \infty$, we find that (6) is roughly proportional to a $\mathrm{Ga}(b + 0.5 \sum_{i=1}^p (a_i - 1), c + \sum_{i=1}^p a_i \log(s_i/a_i))$. This gives a straightforward manner to obtain approximate draws from (6). To approximate $C(\mathbf{a}, b, c, d, r, \mathbf{s})$, one could simply use the normalization constant of the approximating gamma distribution. Instead, we use an alternative approach that in practice improves the quality of the approximation. Denote as $g(y)$ the probability density function of the gamma approximation, and let $m$ be its mode. Evaluating $g$ and (6) at $m$ gives

$$
\begin{aligned}
C(\mathbf{a}, b, c, d, r, \mathbf{s}) & \\
& \approx g(m) m^{-b+pd+1} e^{mc} \prod_{i=1}^p \frac{\Gamma(m)^{a_i}}{\Gamma(a_i m + d)} \left(\frac{m}{r + s_i m}\right)^{-(a_i m + d)}.
\end{aligned}
\tag{7}
$$

In the Supplementary Material [Rossell (2009)] we show examples comparing the gamma shape distribution and its gamma approximation. In our examples the approximation error is below $10^{-5}$ for the density and $10^{-14}$ for $C(\mathbf{a}, b, c, d, r, \mathbf{s})$. In the microarray data that we have analyzed so far the approximation worked well, as most coefficients of variation are $\ll 1$ and, therefore, the posterior distribution is centered around large $y$ values. Also, $a_i \geq 1$ as it is the sample size in group $i$ and $d > 0$, and, hence, $a_i y + d$ is also large. In some rare cases we detected that the mode of the approximation did not match that of (6) (indicated by the first derivative of $\log[f(y|a, b, c, d, r, s)]$ not being close to zero). In these cases we used a few Newton–Raphson steps to locate the mode and used the gamma approximation that matches the location of the mode as well as the value of the second derivative of $\log[f(y|a, b, c, d, r, s)]$ evaluated at the mode.

**5. Inference.** We formalize inference for differential expression by minimizing the Bayesian false negative rate (BFNR) subject to an upper bound on the Bayesian false discovery rate (BFDR) [Müller, Parmigiani and Rice (2007)]. Briefly, BFNR is the posterior expected false negative rate (i.e., genes declared EE that are actually DE), and BFDR is the posterior expected false positive rate (i.e., genes declared DE that are actually EE). The definition is analogous to the frequentist FDR and FNR definitions, and it remains valid for more than two hypotheses. The Bayes rule is to declare a gene as DE whenever its posterior probability of DE is above a certain threshold. Müller, Parmigiani and Rice (2007) provided a simple expression to determine the threshold. The result extends trivially to our multiple hypotheses setup with a slight adjustment: given that a gene is not classified into pattern 0, we propose assigning it to the pattern with the highest posterior probability. That is, for given BFDR and BFNR we maximize the number of genes correctly classified into their expression pattern.

Since the posterior probabilities in Section 2.2 are derived under an assumed probability model, deviations from the model assumptions may result in poor performance of the procedure. In Section 6.2 we assess its frequentist operating characteristics by bootstrapping one example. For further details see the supplementary material [Rossell (2009)].

**6. Results.** We assess the performance of the GaGa and MiGaGa models in simulated data and several examples. We fit MiGaGa with $M=2$ clusters, as we believe it offers a reasonable compromise between model flexibility and computational speed. In Section 6.1 we analyze the leukemia data of Armstrong et al. (2002), and in Section 6.2 we use this data to conduct several simulation studies. We also analyze the Affymetrix data from the MAQC study [MAQCconsortium (2006)]. This study is a valuable resource, as the differential expression status of over 1,000 genes was validated via quantitative PCR Taqman assays. The models were fit via an EM scheme, as described in Section 2.3.

We compare our methodology to the Ga model, BRIDGE [Gottardo et al. (2006)], limma with Benjamini–Hochberg $p$-value adjustment (limma-BH) [Smyth (2005); Benjamini and Hochberg (1995)], the Significance Analysis of Microarrays (SAM) of Tusher, Tibshirani and Chu (2001), and a $t$-test/$F$-test with beta-uniform mixture $p$-value adjustment ($t$-BUM) [Pounds and Morris (2003)]. For limma-BH, SAM and $t$-BUM we use log2-transformed data BRIDGE to automatically log-transform the data, so we gave un-logged data as input to the routine. For Ga, GaGa and MiGaGa we use the original scale, as these methods have been designed to work with positive real values. All methods are used as implemented in their respective Bioconductor packages [Gentleman et al. (2004)]. Ga is one of the methods implemented in the package EBarrays [Kendziorski, Newton and Sarkar (2005)]. We do not perform a comparison with the lnnmv procedure within EBarrays, which formulates a log-normal model that allows genes to have different variances in a manner similar to limma. BRIDGE is implemented in bridge [Gottardo (2004)], limma-BH in limma [Smyth (2005)], SAM in siggenes [Schwender (2007)], and $t$-BUM in ClassComparison [Coombes (2007)].

All methods were set up to control the FDR below 0.05.

6.1. *Armstrong data.* The data (http://www.broad.mit.edu/cgi-bin/cancer/publications/pub_paper.cgi?mode=view&paper_id=63) consists of 24 Affymetrix U95A arrays from acute lymphoblastic leukemia (ALL) samples, 18 U95A arrays from lymphoblastic leukemia with MLL translocations (MLL), and 2 U95Av2 arrays also from the MLL group. The U95Av2 arrays were obtained at a later date than the rest, possibly under different experimental conditions, so we excluded them from the analysis. The data also contained samples with acute myelogenous leukemia, but for illustration we restrict attention to the ALL and MLL groups.

The data was background corrected, normalized and summarized using the function just.gcrma from the R package gcrma [Wu and Irizarry (2007)].

6.1.1. *Model fit.* Figure 1(a) reveals a violation of the constant CV assumption of the Ga model, and that the model tends to flag genes with large CVs as DE. Figure 1(b) shows that a MiGaGa fit with $M = 2$ components describes the data better than a single-component GaGa. Further assessment of goodness of fit can be found in Supplementary Material [Rossell (2009)].

6.1.2. *Differential expression analysis.* To study the behavior of the methods under small sample sizes and evaluate the reliability of the results, we analyze multiple random subsets of chips and report averaged results. We start by fitting the model to 5 randomly chosen arrays from each group. We then add 5 more arrays per group, then 10 and finally we analyze the full data set. We repeat this process 20 times.

Table 1 shows the average number of genes declared DE when analyzing a subset of 5, 10 and 15 arrays per group, as well as the full data. The table also provides the percentage of reproducibility, that is, how many among those genes were found again when analyzing the full data set. For instance, with 5 arrays per group MiGaGa found 61.5 genes on the average, 86.0% of which were confirmed in the full data. Ga was the method declaring the most genes as DE. The observed lack of fit of the Ga model and the simulation study conducted in Section 6.2 suggest that Ga produces lists of DE genes with an FDR well above the desired 5%. For most sample sizes, GaGa and MiGaGa found more genes than the remaining competitors, and presented reasonably high reproducibility. The advantage is especially noticeable for small sample sizes, where GaGa and MiGaGa find at least twice as many genes as the competitors.

Table 1

*Gene discoveries in the Armstrong data. # DE: average number of genes declared DE; % rep.: average percentage of # DE also found when analyzing the full data*

|  | 5 arrays | | 10 arrays | | 15 arrays | | All data |
|---|---|---|---|---|---|---|---|
|  | # DE | % rep. | # DE | % rep. | # DE | % rep. | # DE |
| GaGa | 58.5 | 0.856 | 431.0 | 0.893 | 784.0 | 0.889 | 991 |
| MiGaGa ($M=2$) | 61.5 | 0.860 | 445.0 | 0.893 | 815.0 | 0.890 | 1040 |
| Ga | 900.0 | 0.810 | 1261.0 | 0.885 | 1526.0 | 0.918 | 1744 |
| BRIDGE | 21.5 | 0.944 | 182.5 | 0.960 | 439.0 | 0.959 | 716 |
| limma-BH | 21.5 | 0.947 | 181.5 | 0.957 | 543.0 | 0.946 | 972 |
| SAM | 0.0 | 0.937 | 274.0 | 0.973 | 804.0 | 0.956 | 1477 |
| $t$-BUM | 7.5 | 1.000 | 168.5 | 0.975 | 586.0 | 0.965 | 1194 |

6.2. *Simulation study.* We conduct a parametric and a nonparametric simulation study. For both we use the Armstrong data, so that the simulations are as realistic as possible. We simulate 200 data sets, conduct analyses analogous to those in Section 6.1 and compute average power, FDR, Receiver Operating Characteristics (ROC) curves and areas under the curve (AUC).

For the parametric simulation, we generate data from the posterior predictive distribution of the GaGa model fit to the Armstrong data. That is, gene expressions are gamma distributed and, for each gene, the simulated means and CVs are consistent with the sample means and CVs. The genes DE status are determined based on the DE posterior probabilities.

For the nonparametric simulation, we also determine which genes are differentially expressed using the GaGa posterior probabilities from Armstrong data. Expression values for EE genes are generated by re-sampling the Armstrong data arrays with replacement, regardless of which group they came from. This conserves both the marginal distribution for each gene and also the correlations between EE genes. For DE genes, we again sample arrays with replacement, but the ALL data is simulated by sampling from the ALL group only, and similarly for MLL data. We repeat this process 200 times and report average power and FDR.

6.2.1. *Differential expression.* Table 2 shows the observed FDR and power for several sample sizes. As suggested by the lack of fit discussed above, Ga presents FDR rates well above the desired 5%. In the parametric simulations, GaGa and MiGaGa appropriately control the FDR below 5% and they present a higher average power than the remaining competitors for all sample sizes. In the nonparametric simulations, the advantage in power is more noticeable, that is, from 0.449 for limma-BH to 0.671 for MiGaGa. However, the FDR was slightly above 5% in several scenarios. Among the competitors, limma-BH and SAM performed best.

6.2.2. *ROC curves.* We computed the average ROC curve over the 200 nonparametric simulations with 20 arrays per group. Figure 2(a) shows the average FDR and power. GaGa and MiGaGa presented very similar curves and dominated uniformly the competing methods. limma-BH, SAM and *t*-BUM performed better than Ga and BRIDGE when the FDR was below 0.2, that is, the range typically considered in practice. The GaGa AUC was significantly lower than the MiGaGa AUC, and significantly higher than the other methods' AUC (Wilcoxon paired test, Bonferroni corrected $P < 0.01$). However, the numerical difference between the GaGa and MiGaGa AUCs is of no practical relevance.

6.3. *MAQC study.* The MicroArray Quality Control (MAQC) project was initiated to assess the reliability and reproducibility of findings obtained from microarray experiments. Expression data was obtained for four titration pools (A, B, C and D) generated from two distinct reference RNA samples, at multiple sites and using several technology platforms. The two RNA samples types were Universal Human Reference RNA (UHRR) from Stratagene and a Human Brain Reference RNA (HBRR) from Ambion. The four pools included the two reference RNA samples and two mixtures: Sample A, 100% UHRR; Sample B, 100% HBRR; Sample C, 75% UHRR:25% HBRR; and Sample D, 25% UHRR:75% HBRR. Confirmatory qPCR assays were performed for the 1,296 genes with the largest $t$-test statistic values comparing groups A vs. B. qPCR assays have a much higher sensitivity than microarrays, and are therefore a powerful tool to assess the DE status of a selected list of genes.

TABLE 2
*Average FDR and power for different sample sizes. Data simulated from GaGa posterior predictive*

|  | 5 arrays | | 10 arrays | | 15 arrays | | 20 arrays | |
| --- | --- | --- | --- | --- | --- | --- | --- | --- |
|  | FDR | Power | FDR | Power | FDR | Power | FDR | Power |
| Parametric simulation | | | | | | | | |
| GaGa | 0.011 | 0.066 | 0.000 | 0.322 | 0.007 | 0.512 | 0.002 | 0.608 |
| MiGaGa ($M = 2$) | 0.011 | 0.066 | 0.000 | 0.328 | 0.008 | 0.520 | 0.002 | 0.615 |
| Ga | 0.159 | 0.434 | 0.133 | 0.587 | 0.117 | 0.667 | 0.105 | 0.712 |
| BRIDGE | 0.000 | 0.034 | 0.016 | 0.232 | 0.032 | 0.424 | 0.022 | 0.553 |
| limma-BH | 0.012 | 0.063 | 0.035 | 0.288 | 0.034 | 0.487 | 0.036 | 0.580 |
| SAM | 0.000 | 0.000 | 0.044 | 0.272 | 0.043 | 0.492 | 0.042 | 0.582 |
| *t*-BUM | 0.065 | 0.021 | 0.040 | 0.266 | 0.034 | 0.480 | 0.042 | 0.577 |
| Nonparametric simulation | | | | | | | | |
| GaGa | 0.043 | 0.054 | 0.066 | 0.319 | 0.067 | 0.529 | 0.065 | 0.660 |
| MiGaGa ($M = 2$) | 0.047 | 0.057 | 0.066 | 0.327 | 0.068 | 0.541 | 0.068 | 0.671 |
| Ga | 0.342 | 0.397 | 0.289 | 0.567 | 0.254 | 0.666 | 0.239 | 0.740 |
| BRIDGE | 0.099 | 0.048 | 0.045 | 0.133 | 0.035 | 0.240 | 0.041 | 0.339 |
| limma-BH | 0.047 | 0.029 | 0.021 | 0.168 | 0.019 | 0.321 | 0.024 | 0.449 |
| SAM | 0.049 | 0.020 | 0.050 | 0.197 | 0.051 | 0.360 | 0.048 | 0.481 |
| *t*-BUM | 0.070 | 0.022 | 0.043 | 0.156 | 0.053 | 0.324 | 0.048 | 0.454 |

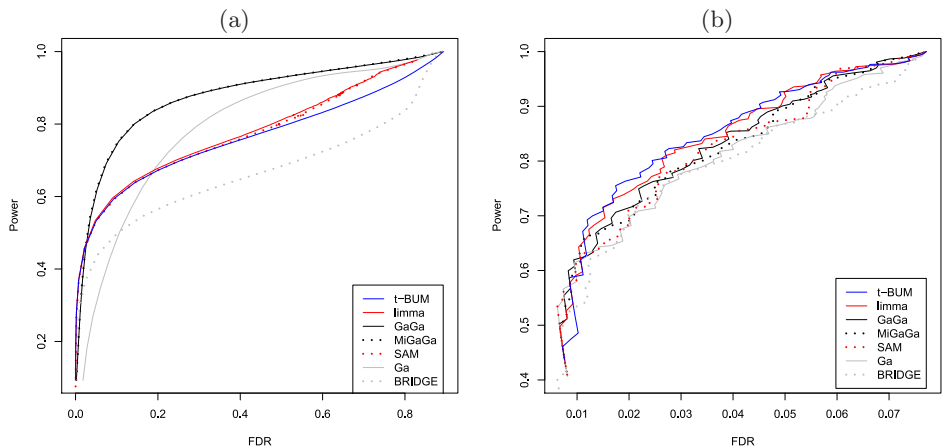

Fig. 2. *ROC curves.* (a) *Averaged over Armstrong bootstrap simulations. AUC: GaGa = 0.781, MiGaGa = 0.782, Ga = 0.707, BRIDGE = 0.595, limma-BH = 0.631, SAM = 0.620, t-BUM = 684.* (b) *MAQC data. AUC: GaGa = 0.0585, MiGaGa = 0.0581, Ga = 0.0570, BRIDGE = 0.0567, limma-BH = 0.0596, SAM = 0.0580, t-BUM = 0.0599.*

We restrict attention to the 20 Affymetrix hgu133plus2 arrays obtained in the first site, and we analyze the microarray data as pre-processed in MAQCconsortium (2006). We consider that qPCR confirms that a gene is DE whenever the limma $F$-test unadjusted $p$-value for the qPCR data is less than 0.05.

6.3.1. *Differential expression.* When fitting the GaGa and MiGaGa ($M = 2$) models we reflect the experimental setup considering the following 5 hypotheses for each gene:

(8)
$$\begin{aligned}
&\text{Pattern 0: A} = \text{C} = \text{D} = \text{B}, \\
&\text{Pattern 1: A} \neq \text{C} = \text{D} = \text{B}, \\
&\text{Pattern 2: A} = \text{C} \neq \text{D} = \text{B}, \\
&\text{Pattern 3: A} = \text{C} = \text{D} \neq \text{B}, \\
&\text{Pattern 4: A} \neq \text{C} \neq \text{D} \neq \text{B}.
\end{aligned}$$

As groups C and D contain both UHRR and HBRR, one expects a priori that most genes follow either Pattern 0 or Pattern 4. Patterns 1–3 include DE genes for which not all titrations are different. For instance, Pattern 1 contains genes for which 25% HBRR is enough to modify its expression levels as much as 100% HBRR.

GaGa assigns 20272, 0, 1429, 3935 and 29039 genes to patterns 0, 1, 2, 3 and 4 (respectively), whereas MiGaGa assigns 16328, 0, 1323, 3697 and 33327. That is, most of the differentially expressed genes are assigned to Pattern 4, as expected. We assessed the biological interpretation of these gene lists by conducting a gene ontology enrichment analysis. To this end, we created (i) a list with the 1000 genes with the highest posterior probability for Pattern 4 and that presented group B mean > group A mean, and (ii) another list with 1000 with highest posterior probabilities for Pattern 4 and group A mean > group B mean. After obtaining these two lists both for GaGa and MiGaGa, we tested for enriched biological process categories using the DAVID software [Dennis et al. (2003)]. The highest enrichment in genes over-expressed in group B with respect to group A was observed in transmission of nerve impulse, synaptic transmission, nervous system development, cell–cell signaling and cell communication. All these functions are needed specifically in the brain, and, hence, they should indeed be enriched in HBRR. The highest enrichment in genes under-expressed in group B was observed for cell cycle, cell cycle phase, mitotic cell cycle, cell cycle process and M phase. These functions are all related to cell proliferation, and are expected to be turned off in differentiated organs like the brain. The complete listings of the significantly enriched categories were also highly consistent with the experiment's biological background, and are provided in the Supplementary Material [Rossell (2009)].

6.3.2. *ROC curves.* We compare our approach to the competing methods by computing ROC curves, using only qPCR-validated genes. It should be noted that, as only genes with large $t$-test statistic values were selected for qPCR validation, this puts the $t$-test based methods (limma-BH, SAM, $t$-BUM) at an advantage with respect to Ga, GaGa, MiGaGa and BRIDGE. That is, a number of the potentially interesting genes found by the latter methods were never validated due to having small to moderate $t$-test statistic values.

For Ga, we specify the same 5 expression patterns considered for GaGa and MiGaGa. For the remaining methods, we simply test the null hypothesis (Pattern 0) against the full alternative (Pattern 4). As the Bioconductor bridge package currently does not support more than 3 groups, for BRIDGE we only analyzed data from groups A and B.

Figure 2(b) displays the results. The $x$-axis provides proportion of non-validated qPCR genes (i.e., with limma-BH $p$-value >0.05 for qPCR data), and the $y$-axis the proportion of validated qPCR genes (i.e., $p$-value >0.05). Among the 1296 genes selected for qPCR validation in the original MAQC paper, only 7.8% were not confirmed. Therefore, for all considered analysis methods the proportion of nonvalidated genes ($x$-axis) can at the most be 7.8%.

In contrast with Figure 2(a), the differences between methods are not large. The $t$-test based methods present slightly higher AUC than GaGa and MiGaGa, which in turn present slightly higher AUC than Ga and BRIDGE. Qualitatively, all methods achieve good levels of qPCR validation.

**7. Discussion.** We introduced two hierarchical models for high-throughput data based on the gamma distribution. This flexible parametric choice allows to capture the asymetric data frequently encountered in this field, even after log-transformations. GaGa builds on the Ga model by relaxing the constant coefficient of variation assumption. This results in a parsimonious model with a substantial improvement in the model fit and therefore in reliability of the resulting inference. The increased generality comes at a negligible computational cost. We derived an approximation for the posterior distribution of the gamma shape parameter that further simplifies computation. The second extension, the MiGaGa, increases the model flexibility by incorporating a mixture prior, at the expense of model parsimony. In practice, a mixture with as few as two components may suffice to provide a satisfactory fit. We have shown that, in many situations, GaGa achieves almost the same degree of performance compared to MiGaGa, and hence that it is a sensible default choice.

The hierarchical nature of the models allows for the sharing of information across genes. This is specially beneficial for the small sample sizes often encountered in high-throughput experiments.

We compared our models with two other gamma-based models and three $t$-test and normal linear model based approaches. In simulations and in several examples we have shown how both GaGa and MiGaGa find more genes than the competing methods, while controlling the FDR around the desired levels. For instance, when analyzing a subset with 5 arrays per group from the Armstrong data we detect around 60 differentially expressed genes, while the best competing methods found around 20. The fact that around 86% of these genes were found again when analyzing the full data gives us confidence that these are not spurious findings. Both in parametric and nonparametric simulations we have seen that GaGa and MiGaGa present improved operating characteristics. ROC curves showed potential for substantial increases in power at fixed FDR levels. Even in a list of qPCR-validated genes selected for having a large $t$-test statistic, our models performed almost as well as the $t$-test based methods, and delivered biologically meaningful results.

Some extensions of the model are possible. For instance, the interest might be not only in seeking differences in mean expression but also in distributional shape. This is frequently of biological interest, since many mutations, deletions and translocations affect only a proportion of the diseased individuals, and, hence, one expects to see differences in the tail behavior between groups. Although not presented in this paper, we derived such an approach and implemented it as an option in the Bioconductor `gaga` library.

Also, we have not explicitly modeled the dependence between genes. Not learning about the dependence structure limits the use of the model in finding gene networks or gene interactions. Interesting future work will be to include dependence. Other possibilities are extending the model to include covariate information and study-specific random effects, which would make it appealing for meta-analysis purposes, or using the model for sample size calculation as in Müller et al. (2004) or sequential sample size calculation. In the latter application, the computational efficiency of the GaGa model should prove a major asset.

**Acknowledgments.** We thank Peter Müller and Jens Luders for their very useful comments.

## SUPPLEMENTARY MATERIAL

**Supplement to GaGa: A parsimonious and flexible model for differential expression analysis** (DOI: 10.1214/09-AOAS244SUPP; .pdf). We detail an EM algorithm and two fully Bayesian MCMC schemes for model fitting, and a Bayesian procedure for FDR control. We also assess model goodness-of-fit, assess the quality of the gamma approximation to the gamma shape distribution and detail the gene ontology analysis performed for the MAQC study.

Baldiri Reixac, 15
Barcelona
Spain 08028
E-mail: rosselldavid@gmail.com
URL: http://rosselldavid.googlepages.com